\documentclass[10pt]{amsart}
\usepackage{amsmath}
\usepackage{latexsym}
\usepackage{amsfonts}
\usepackage{amssymb}
\usepackage{amsthm,amsxtra}
\usepackage{nicefrac}
\usepackage{portland}
\usepackage{rotating}

\theoremstyle{plain}
\newtheorem{theorem}{Theorem}

\newtheorem{lemma}{Lemma}
\newtheorem{proposition}{Proposition}

\theoremstyle{definition}

\theoremstyle{remark}
\newtheorem{remark}{Remark}

\newcommand{\Z}{\mathbb Z}

\newcommand{\C}{\mathbb C}
\newcommand{\+}{\!+\!}
\newcommand{\m}{\!-\!}
\newcommand{\PV}{${\rm P}_{\rm V}\;$}
\newcommand{\PVI}{${\rm P}_{\rm VI}\;$}
\newcommand{\PIII}{${\rm P}_{\rm III}\;$}
\newcommand{\III}{\rm III}
\newcommand{\PIIIdash}{${\rm P}_{\rm III^{\prime}}\;$}

\newcommand{\half}{\nicefrac{1}{2}}
\newcommand{\quarter}{\nicefrac{1}{4}}
\newcommand{\eight}{\nicefrac{1}{8}}
\newcommand{\thirtytwo}{\nicefrac{1}{32}}
\newcommand{\threehalf}{\nicefrac{3}{2}}

\begin{document}
\noindent
{\bf
Gap Probabilities for Double Intervals in Hermitian Random Matrix Ensembles as $\tau$-Functions
-- Spectrum Singularity case}

\vspace{5mm}
\noindent
N.S.~Witte\\

\noindent
Department of Mathematics and Statistics
and School of Physics, \\
University of Melbourne, Victoria 3010, Australia \\
Email: {\tt N.Witte@ms.unimelb.edu.au}

\begin{quote}
The probability for the exclusion of eigenvalues from an interval $(-x,x)$ 
symmetrical about the origin for a scaled ensemble of Hermitian random matrices, 
where the Fredholm kernel is a type of Bessel kernel with parameter $ a $ 
(a generalisation of the sine kernel in the bulk scaling case), 
is considered. It is shown that this probability is the square of a $\tau$-function, 
in the sense of Okamoto, for the Painlev\'e system \PIII. This then leads to a 
factorisation of the probability as the product of two $\tau$-functions for the 
Painlev\'e system \PIIIdash. A previous study has given a formula of this type 
but involving \PIIIdash systems with different parameters consequently implying 
an identity between products of $\tau$-functions or equivalently sums of Hamiltonians.
\end{quote}

The probability $ E_{\beta}(0;J;g(x);N) $ that a subset of the real line $ J $ 
is free of eigenvalues for an ensemble of $ N\times N $ random matrices with 
eigenvalue probability density function proportional to
\begin{equation}
   \prod^N_{l=1}g(x_l) \prod_{1 \leq j<k \leq N}|x_j-x_k|^{\beta} ,
\end{equation}
($ \beta=1,2 $ or $ 4 $ according to the ensemble exhibiting orthogonal, unitary 
or symplectic symmetry respectively) is a fundamental statistic in the study of 
these ensembles. Most effort has focused on the case where 
$ J $ is a single interval, one endpoint fixed at the edge of the support
of the measure defining the ensemble whilst the other is free, and taken to
be the independent variable in the system of equations determining the
gap probability. There is, however, another interesting case where the set
$ J $ consists of two disconnected intervals, but related to each other so
that there is still only one free variable. For instance there is the result
for unitary ensembles of Hermitian matrices that the gap probability for an 
interval symmetrical about the origin $ J $ and an even weight function 
$ g_2(x) $ (where the integrable examples include Gaussian, symmetric Jacobi 
and Cauchy weights) factorises \cite{Fo_1999}
\begin{multline}
  E_2(0;J;g_2(x);N) \\
  = E_2(0;J^{+};y^{-1/2}g_2(y^{1/2});\lfloor (N\+1)/2 \rfloor)
    E_2(0;J^{+};y^{ 1/2}g_2(y^{1/2});\lfloor N/2 \rfloor) \ ,
\label{E2_factorise}
\end{multline}
($ J^{+} $ is the positive member of a pair of intervals composing $ J $).
Examples of where this relation has been useful can be found in the above
reference and \cite{FW_2002c}.
An example of such a double interval statistic is one arising from an ensemble 
of random unitary $ N\times N $ matrices with the joint eigenvalue pdf
\begin{equation}
   p(z_1,\ldots,z_N) = 
   C_{N,a}\prod^N_{l=1}|1-z_l|^{2a}\prod_{1 \leq j<k \leq N}|z_j-z_k|^2,
   \qquad z_l=e^{i\theta_l} .
\label{SS_pdf}
\end{equation}
The weight function has an algebraic singularity at $ z = e^{i\theta} = 1 $
corresponding in the log-gas picture of (\ref{SS_pdf}) to an impurity charge of 
$ a $, so the ensemble is termed the {\it spectrum singularity} case. The 
probability that no eigenvalues $ e^{i\theta} $ have phases in the interval 
$ (-\theta,\theta) $ is given by
\begin{multline}
  E_2(0;(-\theta,\theta);|1-z|^{2a};N) \\
  = \int_{(-\pi,-\theta)\cup(\theta,\pi)}\frac{dz_1}{2\pi iz_1}
    \ldots
    \int_{(-\pi,-\theta)\cup(\theta,\pi)}\frac{dz_N}{2\pi iz_N}
    p(z_1,\ldots,z_N) .
\end{multline}
This problem can also be equivalently expressed in terms of a problem with
its spectrum on the real line, where the weight function is now a Cauchy
one, and the above gap probability is
$ E_2(0;(-\infty,-s)\cup(s,\infty);(1+\lambda^2)^{-N-a};N) $.

We wish to focus here on the specific example of a double interval 
$ J=(-x,x) $ for the bulk scaling limit of the Spectrum Singularity case, 
as $ N \to \infty $. The gap probability is known to be expressed as a
Fredholm determinant
\begin{equation}
   E^{\rm SS}_2(0;(-x,x);a) = \det(1-{\mathbb K}_J) ,
\end{equation}
where the integral operator $ {\mathbb K} $ has the kernel
\begin{equation}
   K(x,y) = \sqrt{\pi x}\sqrt{\pi y}\frac{
   J_{a+1/2}(\pi x)J_{a-1/2}(\pi y)
  -J_{a+1/2}(\pi y)J_{a-1/2}(\pi x)}{2(x-y)} ,
\label{SS_kernel}
\end{equation}
with a parameter $ a \in \C $ with $ {\rm Re}(a) > -1/2 $ and the density 
of eigenvalues is $ \rho = 1 $. 
The Tracy and Widom theory for the Fredholm determinant forms of gap 
probabilities \cite{TW_1994} was employed in \cite{FO_1996,WF_2000}, and this 
probability was evaluated as
\begin{equation}
   E^{\rm SS}_2(0;(-x,x);a) = \exp\left( \int^{2\pi x}_0 dy\,
           \frac{\sigma_1(y)}{y} \right) .
\label{SS_gap}
\end{equation}
in terms of $ \sigma_1(r) \equiv -2xR(x,x) $ with $ r = 2x $, which was shown 
to satisfy the ordinary differential equation
\begin{equation}
   (r\sigma_1'')^2 + 4[-a^2-\sigma_1+r\sigma_1']
    \left\{ (\sigma_1')^2
            - \left[ a-\sqrt{a^2+\sigma_1-r\sigma_1'} \right]^2
    \right\} = 0 ,
\label{SS_ode}
\end{equation}
subject to the boundary condition
\begin{multline}
   \sigma_1(r) \mathop{\sim}\limits_{r \to 0^+}
   \\
    C_a r^{2a+1} \left( 1-\frac{a}{2(2a+3)(2a+1)}r^2
     +\frac{a}{16(2a+5)(2a+3)(2a+1)}r^4 + \ldots \right)
   \\
    -C^2_a \frac{r^{4a+2}}{2a+1} \left( 1-\frac{a+1}{(2a+3)^2}r^2 + \ldots 
     \right) 
    +C^3_a \frac{r^{6a+3}}{(2a+1)^2} \left( 1 + \ldots \right)
    + \ldots ,
\label{SS_bc}
\end{multline}
where
\begin{equation}
   C_a = -\frac{2}{4^{2a+1}\Gamma(a\+\half)\Gamma(a\+\threehalf)} .
\end{equation}
When $ a=0 $ this equation reduces to case of the bulk scaling limit first
found in \cite{JMMS_1980}, and is then a special case of the Jimbo-Miwa-Okamoto
$\sigma$-function form of \PV. 
Equation (\ref{SS_ode}) was solved in terms of Painlev\'e's fifth transcendent
\cite{WF_2000} for general $ a $ with the parameters
\begin{equation}
  \alpha=\thirtytwo(1-2a)^2,\; \beta=-\thirtytwo(1-2a)^2,\;
  \gamma=0,\; \delta=-2 \ .
\label{SS_PV}
\end{equation}

However there is a puzzling aspect to this result, which also 
appears in the cases of double intervals with other weights such as the 
Gaussian, symmetric Jacobi \cite{WFC_2000} or Cauchy \cite{WF_2000} ones.
Expressed as a second-order second-degree ODE (\ref{SS_ode}) has radical terms, 
namely the square-root term, or alternatively is a quartic second-order ordinary
differential equation in polynomial form. 
Consequently, the ordinary differential equation (\ref{SS_ode}) is not of the 
Jimbo-Miwa-Okamoto $\sigma$-form of \PV for $ a \neq 0 $, as is the case for 
the single intervals. On this point we remark there is now a large body of works 
\cite{FW_2001a,FW_2002a,FW_2002b} demonstrating that gap probabilities and their
generalised averages for unitary random matrix ensembles with classical weights
can be evaluated in terms of a single $\tau$-function for one of the Painlev\'e
systems. Consequently the logarithmic derivatives of these averages satisfy
a Jimbo-Miwa-Okamoto $\sigma$-form for the appropriate system, which are 
generically only second-order second-degree ordinary differential equations
(see the classification of such ODEs with the Painlev\'e property by Cosgrove 
and Scoufis \cite{CS_1993}).
Therefore it was not clear how the spectrum singularity result fitted into this
broader scheme. We provide an answer to this question in this work - there is a
natural explanation in the Okamoto theory for the Painlev\'e transcendents
\PIII and \PIIIdash.

Firstly we recognise that the special case (\ref{SS_PV}) of \PV is one that 
degenerates to \PIII using the transformations of Gromak \cite{Gr_1975a,Gr_1984} 
- for example this can be achieved with the parameters 
$ (v_1,v_2) = (a-\half,a-\half) $ or 
$ \alpha_{\rm III}=1-2a, \beta_{\rm III}=1+2a, \gamma_{\rm III}=1, \delta_{\rm III}=-1 $.
Equation (\ref{SS_ode}) arises quite naturally in the Painlev\'e III system, as will be 
apparent from Okamoto's theory for \PIII \cite{Ok_1987c}. 
The \PIII differential equation is
\begin{equation}
  q'' = \frac{1}{q}\left(q'\right)^2
       - \frac{1}{t}q'
       + \frac{1}{t}(\alpha q^2+\beta)
       + \gamma q^3
       + \frac{\delta}{q} ,
\end{equation}
and can be generated from the Hamiltonian system
\begin{equation}
  tH_{\III} =
  2q^2p^2 - \left[ 2\eta_{\infty}tq^2+(2v_1+1)q-2\eta_{0}t \right]p
          + \eta_{\infty}(v_1+v_2)tq .
\label{PIII_ham}
\end{equation}
One then recovers the standard \PIII ODE for $ q(t) $ with the parameter
identifications
\begin{equation}
  \alpha = -4\eta_{\infty}v_2,\; \beta = 4\eta_{0}(v_1+1),\;
  \gamma = 4\eta^2_{\infty},\; \delta = -4\eta^2_{0} .
\end{equation}
Here $ \eta_{\infty}, \eta_{0} $ are arbitrary parameters which control the
scaling of the independent and dependent variables, so they are usually fixed
at some nominal value (unity). Now let us define the auxiliary Hamiltonian
\begin{equation}
  h = tH_{\III} + \eight(2v_1+1)^2 ,
\label{PIII_aux}
\end{equation}
and examine the time evolution of this. 

\begin{theorem}[Proposition 1.9, \cite{Ok_1987c}]
The auxiliary Hamiltonian $ h(t) $ for \PIII, as specified by (\ref{PIII_aux}) 
with parameters $ v_1, v_2 $ satisfies the ordinary differential equation
\begin{multline}
   (th'')^2 = \left[ 2(h-th') \right] \Bigl\{
   4(h')^2 + 16\eta_{0}\eta_{\infty}\left[ 2(h-th') \right]
   \\
           - 16\eta_{0}\eta_{\infty}\epsilon (v_2-v_1-1) \sqrt{2(h-th')}
           - 16\eta_{0}\eta_{\infty}(v_2-\half)(v_1+\half)
             \Bigr\} ,
\label{PIII_AHode}
\end{multline}
with an arbitrary sign $ \epsilon = \pm 1 $.
\end{theorem}
\noindent Proof:
One can verify that the canonical variables are given by
\begin{align*}
   4\eta_{0}p & =
   h' - \epsilon\frac{th''}{\sqrt{ 8(h-th') }} \\
   2\eta_{\infty}q & =
   \frac{ h' + \epsilon\dfrac{th''}{\sqrt{ 8(h-th') }}}
        { v_2-\half-\epsilon\sqrt{ 2(h-th') } } ,
\end{align*}
with $ \epsilon = \pm 1 $ provided that $ h(t) $ is not a singular solution of
(\ref{PIII_AHode}). Employing these two relations in
\begin{equation*}
   8(h-th') = \left( 4qp -2v_1 -1 \right)^2 ,
\end{equation*}
one arrives at the stated ordinary differential equation (\ref{PIII_AHode}).
\hfill$\square$

\begin{remark}
Okamoto gives a variant of this (modulo typographical mistakes) for
$ \epsilon = 1 $ where it is rearranged and squared to render it
polynomial. 
\end{remark}

\begin{remark}
A more restricted version of this can be found using the Ablowitz-Fokas method 
of Riccati transformations \cite{FA_1982}. Thus by extending their transformation
(Theorem 4.3) relating the solution $ v(z) $ of a \PIII equation with general
parameters to that of a second-order quadratic ODE with solution $ \phi(t) $, 
then defining a new variable $ w(t) $ by
\begin{equation*}
   \phi^2 = 4(tw'-w)+\eta \ ,
\end{equation*}
one finds $ w(t) $ satisfies a particular form of (\ref{PIII_AHode}).
\end{remark}

By making scale changes to dependent and independent variables (\ref{SS_ode}) can 
be brought into correspondence to (\ref{PIII_AHode}) and one solves for the
parameters,
\begin{gather}
   v_1 = -\epsilon a-\half, \qquad
   v_2 =  \epsilon a+\half
   \label{SS_param:a} \\
   \eta_{0} = \eta_{\infty} = 1, \qquad r = 4it
   \label{SS_param:b} \\
   \sigma_1(r) = 2h(t)-a^2 = 2tH_{\III}(t) .
   \label{SS_param:c}
\end{gather}
Thus $ \sigma_1(r)/r $ is proportional to the Hamiltonian for the \PIII system,
with parameters as specified. Furthermore, introducing the $ \tau$-function 
$ \tau_{\rm III}(t) $ for the \PIII system by the requirement
\begin{equation}
   H_{\rm III}(t) =: \frac{d}{dt}\log\tau_{\rm III}(t) ,
\end{equation}
it follows from (\ref{SS_param:b}), (\ref{SS_param:c}) and (\ref{SS_gap}) that
\begin{equation}
   E^{\rm SS}_2(0;(-x,x);a) = 
   \tau^2_{\rm III}(\frac{\pi x}{2i}) \Big|_{v_1=-v_2=-\epsilon a-\half} .
\end{equation}

So far everything is in terms of the \PIII system, but now we will find that 
conversion to the \PIIIdash system yields the desired $\tau$-function
representation. The \PIIIdash system $\{ s, q_{\III'}, p_{\III'}, H_{\III'} \}$
is entirely equivalent to the \PIII system but differs in a number of salient
features, one of which is that the transcendent $ q_{\III'}(s) $ has
movable double poles in contrast to the single poles for $ q_{\III}(t) $.

The Hamiltonian for the \PIIIdash system is given by \cite{Ok_1987c}
\begin{equation}
   sH_{\rm III'}(s) = q_{\III'}^2p_{\III'}^2
   -(q_{\III'}^2+v_1q_{\III'}-s)p_{\III'} + \half(v_1+v_2)q_{\III'} ,
\end{equation}
and the associated $\tau$-function is given by
\begin{equation}
   H_{\rm III'} =: \frac{d}{ds}\log\tau_{\rm III'}(s) .
\end{equation}
From the Hamiltonian the $\sigma$-function $ \sigma_{\rm III'} $ is defined by
\begin{equation}
  \sigma_{\rm III'}(s) = -(sH_{\rm III'})(s/4)-\frac{1}{4}v_1(v_1-v_2)+\frac{1}{4}s . 
\end{equation}
The function $ \sigma_{\rm III'}(s) $ for general $ {\bf v}=(v_1,v_2) $ satisfies
the $\sigma$-form second-order second-degree ordinary differential equation
\begin{equation}
  (s\sigma_{\rm III'}'')^2 
  -v_1v_2(\sigma_{\rm III'}')^2
  +\sigma_{\rm III'}'(4\sigma_{\rm III'}'-1)(\sigma_{\rm III'}-s\sigma_{\rm III'}')
  -\frac{1}{4^3}(v_1-v_2)^2 = 0 .
\label{PIIIdash_sigma}
\end{equation}

\begin{lemma}
The Hamiltonian for the \PIII system is related to that of the \PIIIdash system
by
\begin{equation}
 tH_{\III}(t)\Big|_{{\bf v}} 
 = sH_{\III'}(s)\Big|_{{\bf v}} + sH_{\III'}(s)\Big|_{T_2({\bf v})} ,
\label{PIII_sum}
\end{equation}
where
\begin{equation}
  T_2({\bf v}) = (v_1\+ 1,v_2\m 1) ,
\label{PIIIdash_T2}
\end{equation}
and 
\begin{equation}
  t^2 = s .  
\end{equation}
\end{lemma}

\begin{proof}
Using the mapping relating the two systems \cite{Ok_1987c}
\begin{align}
   tq_{\III}(t) & = q_{\III'}(s) \\
   \frac{1}{t}p_{\III}(t) & = p_{\III'}(s) \\
   tH_{\III}(t) & = 2sH_{\III'}(s) - q_{\III'}(s)p_{\III'}(s) ,
\end{align}
we notice that the right-hand side of the last equality can be written
\begin{align}
   2sH_{\III'}(s) - q_{\III'}p_{\III'} & =
   sH_{\III'}(s) + \left( sH_{\III'}(s) - q_{\III'}p_{\III'} \right)
   \\
   & = sH_{\III'}(s) + T_2( sH_{\III'}(s) )
   \\
   & = sH_{\III'}\Big|_{(v_1,v_2)} + sH_{\III'}\Big|_{(v_1+1,v_2-1)} ,
\end{align}
where the Schlesinger transformation $ T_2 $ acts on the parameters according to
(\ref{PIIIdash_T2}) and upon the Hamiltonian according to
$ T_2(sH_{\III'}) = \left. sH_{\III'}\right|_{{\bf v} \to T_2({\bf v})} $
(see Table \ref{t2}).
Thus we have (\ref{PIII_sum}).
\end{proof}

\begin{proposition}
The gap probability (\ref{SS_gap}) is the product of two $ \tau$-functions for 
the system \PIIIdash
\begin{equation}
   E^{\rm SS}_2(0;(-x,x);a) =
   \tau_{\III'}(-\quarter(\pi x)^2)
                \Bigl|_{(-\epsilon a-\half, \epsilon a+\half)}
   \tau_{\III'}(-\quarter(\pi x)^2)
                \Bigl|_{(-\epsilon a+\half, \epsilon a-\half)} ,
\label{SS_product}
\end{equation}
for $ \epsilon = \pm 1$.
The relevant solutions to (\ref{PIIIdash_sigma}) appearing here are those with
$ {\bf v} = (\mu,-\mu) $ and $ \mu = -\epsilon a-\half $ or $ \mu = -\epsilon a+\half $
satisfying the boundary condition
\begin{multline}
  \sigma_{\rm III'}(s) \mathop{\sim}\limits_{s \to 0^-}
  -\frac{\mu^2}{2}+\frac{s}{4}
  \\
  + \tilde{C}_{\mu}(-s)^{1-\mu}
    \left( 1-\frac{1}{2(\mu-2)}s+\frac{2\mu-3}{16(\mu-3)(\mu-2)(\mu-1)}s^2+\ldots \right)
  \\
  - \tilde{C}^2_{\mu}\frac{(-s)^{2-2\mu}}{\mu-1} \left( 1-\frac{2\mu-3}{2(\mu-2)^2}s+\ldots \right)
  + \tilde{C}^3_{\mu}\frac{(-s)^{3-3\mu}}{(\mu-1)^2} \left( 1+\ldots \right)
  + \ldots ,
\end{multline}
where 
\begin{equation}
  \tilde{C}_{\mu} = \frac{1}{4^{-\mu+1}\Gamma(-\mu+2)\Gamma(-\mu+1)} .
\end{equation}
\end{proposition}

There is another known product form of (\ref{SS_gap}) \cite{Fo_1999},\cite{rmt_Fo}
which states
\begin{equation}
   E^{\rm SS}_2(0;(-x,x);a) =
   E^{\rm HE}_2(0;(0,\pi^2x^2);a-\half)
   E^{\rm HE}_2(0;(0,\pi^2x^2);a+\half) ,
\label{SS_HEproduct}
\end{equation}
where the hard edge gap probability has the evaluation
\begin{equation}
  E^{\rm HE}_2(0;(0,X);a) 
  = e^{-X/4}\tau_{\III'}(\quarter X)\Bigl|_{{\bf v}=(a,a)}
  = \exp\left(-\int^X_0 \frac{ds}{s}\sigma_{\rm III'}(s) \right)
                \Bigl|_{{\bf v}=(a,a)} ,
\end{equation}
again in terms of $\tau$-functions for the \PIIIdash system.
The particular solution of (\ref{PIIIdash_sigma}) arising in the hard edge case 
satisfies the boundary condition
\begin{multline}
  \sigma_{\rm III'}(s) \mathop{\sim}\limits_{s \to 0^+}
   C_as^{a+1} \left( 1-\frac{1}{2(a+2)}s+\frac{2a+3}{16(a+3)(a+2)(a+1)}s^2+\ldots \right)
  \\
  + C^2_a\frac{s^{2a+2}}{a+1} \left( 1-\frac{2a+3}{2(a+2)^2}s+\ldots \right)
  + C^3_a\frac{s^{3a+3}}{(a+1)^2} \left( 1+\ldots \right)
  + \ldots ,
\end{multline}
where
\begin{equation}
  C_a = \frac{1}{2^{2a+2}\Gamma(a+2)\Gamma(a+1)} .
\end{equation}
Combining these two product forms we have the general identity
\begin{multline}
    e^{-2X}\tau_{\III'}(X)\Bigl|_{(a-\half,a-\half)}
            \tau_{\III'}(X)\Bigl|_{(a+\half,a+\half)}
   \\
 = \tau_{\III'}(-X)\Bigl|_{(-\epsilon a-\half, \epsilon a+\half)}
   \tau_{\III'}(-X)\Bigl|_{(-\epsilon a+\half, \epsilon a-\half)} , 
\label{PIII_tauId}
\end{multline}
or an additive relation in terms of the Hamiltonian functions,
\begin{multline}
    -2s+sH_{\III'}(s)\Bigl|_{(a-\half,a-\half)}
       +sH_{\III'}(s)\Bigl|_{(a+\half,a+\half)}
   \\
 = sH_{\III'}(-s)\Bigl|_{(-\epsilon a-\half, \epsilon a+\half)}
  +sH_{\III'}(-s)\Bigl|_{(-\epsilon a+\half, \epsilon a-\half)} .
\label{PIII_HamId}
\end{multline}

A direct derivation of the latter relation can be deduced from the actions of the 
reflection operators $ s_2 $ and $ s_{-}s_2 $ which have the actions 
$ s_2: v_2 \mapsto -v_2, s \mapsto -s $ and $ s_{-}s_2: v_1 \mapsto -v_1, s \mapsto -s $.
Thus from Table 1 we see that
\begin{gather}
 sH_{\III'}(-s)\Bigl|_{(-v_1,v_2)}
 = s_{-}s_2 sH_{\III'}(s)\Bigl|_{(v_1,v_2)} = sH_{\III'}(s)\Bigl|_{(v_1,v_2)}-s
 \\
 sH_{\III'}(-s)\Bigl|_{(v_1,-v_2)}
 = s_2 sH_{\III'}(s)\Bigl|_{(v_1,v_2)} = sH_{\III'}(s)\Bigl|_{(v_1,v_2)}-s ,
\end{gather}
and (\ref{PIII_HamId}) follows by adding two instances of either of the above
two relations.

There is an interesting special case when the $\tau$-functions are classical
solutions and this was found to occur in the studies \cite{Ok_1987c},\cite{FW_2002a}
for $ a \in \Z_{\geq 0}+\half $. For the $\tau$-functions
with parameters on the diagonal ($ a=n-\half, n \in \Z_{>0}  $)
\begin{equation}
  \tau_{\III'}(X)\Bigl|_{(a+\half,a+\half)} 
  = \det \left[ I_{j-k}(2\sqrt{X}) \right]_{j,k=0, \ldots ,n-1} ,
\end{equation}
whereas on the cross-diagonal
\begin{equation}
  \tau_{\III'}(X)\Bigl|_{(-a-\half,a+\half)} 
  = e^{X}\det \left[ J_{j-k}(2\sqrt{X}) \right]_{j,k=0, \ldots ,n-1} ,
\end{equation}
where $ J_{\nu}(z), I_{\nu}(z) $ are the Bessel function and modified Bessel 
function respectively.
Naturally the identity (\ref{PIII_tauId}) is satisfied for these two sets of 
classical solution.

In conclusion our results raise the natural question whether a similar evaluation 
in terms of $\tau$-function products as in (\ref{SS_product}) should exist for the 
gap probabilities defined on symmetric double intervals with the finite $ N $ 
ensembles of Gaussian, symmetric Jacobi and Cauchy weights. These are known to be 
solved in terms of \PV and \PVI transcendents and there are analogues of the 
higher degree ODE (\ref{SS_ode}). Moreover the analogue of the product formula 
(\ref{SS_HEproduct}) is known for each of these cases \cite{Fo_1999}, \cite{rmt_Fo}. 
The next question is whether there is a corresponding Hamiltonian theory underlying 
these ODEs and how it relates to the known Hamiltonian theory. Isolated examples 
of higher degree second-order ODEs (quadratic and quartic in the second derivative) 
are known in the literature through many differing approaches \cite{SM_1997},
\cite{MS_1999},\cite{GP_2001},\cite{Sk_2001}, and one may suspect that these too 
appear in a random matrix context.

\medskip
\noindent{\it Acknowledgements}\\
The author wishes to acknowledge the many and wide-ranging discussions with 
Peter Forrester and the opportunity to visit Chris Cosgrove.
This research has been supported by the Australian Research Council.

\bibliographystyle{cmp}                                                 
\bibliography{random_matrices,nonlinear}
\vfill\eject

\thispagestyle{empty}
\renewcommand{\arraystretch}{2.0}
\begin{table}
\begin{center}
\begin{sideways}
\begin{tabular}{|c||c|c|c|c|c|c|}\hline
 & $v_1$ & $v_2$ & $p$ & $q$ & $s$ & $sH$
\\ \hline
 $s_0$ & $-1-v_2$ & $-1-v_1$  &
 $ \dfrac{q}{t}
   \left[q(p-1) - \frac{1}{2}(v_1-v_2)\right] + 1$ &
 $-\dfrac{t}{q} $ & $s$ & $sH-q(p-1)+\half(v_1-v_2)(1+\half(v_1+v_2))$
\\
 $s_1$ & $v_2$ & $v_1$ & $p$ & 
 $q + \dfrac{v_2-v_1}{2(p-1)} $ & $s$ & $sH-\quarter(v^2_2-v^2_1)$
\\
 $s_2$ & $v_1$  & $-v_2$ &  $1-p$ & $-q$ & $-s$ & $sH-s$
\\
 $s_{-}$ & $-v_1$  & $-v_2$ &  $p$ &
 $q - \dfrac{v_1p-\half(v_1+v_2)}{p(p-1)} $ & $s$ & $sH$
\\
 $ T_2 $ & $v_1+1$ & $v_2-1$ & $\dfrac{q}{t}\left[\half(v_1+v_2)-qp\right]$ &
 $ \dfrac{t}{q}-\dfrac{\half(2+v_1-v_2)t}{q[qp-\half(v_1+v_2)]+t} $ &
 $ s $ & $ sH-qp $
\\ \hline
\end{tabular}
\end{sideways}
\end{center}
\caption{\label{t2}  Generators and selected elements of the group of B\"acklund 
transformations for the \PIIIdash system, the extended affine Weyl group for the root system
$ B^{(1)}_2 $.}
\end{table}
\renewcommand{\arraystretch}{1.0}

\end{document}